\documentclass[aps,prl,floatfix,superscriptaddress,twocolumn,footinbib]{revtex4-2}

\usepackage{array, rotating, booktabs}
\usepackage{diagbox}
\usepackage{siunitx}
\usepackage{float}
\usepackage{amssymb}
\usepackage{amsmath}
\usepackage{amsfonts}
\usepackage{appendix}
\usepackage{bm}
\usepackage{braket}
\usepackage{graphicx}
\usepackage{epsfig}
\usepackage{epstopdf}
\usepackage{balance}
\usepackage[dvipsnames]{xcolor}
\usepackage{calc}
\usepackage{natbib}
\usepackage[colorlinks,
            linkcolor=blue,
            anchorcolor=blue,
            citecolor=blue,
            urlcolor=blue]{hyperref}
\usepackage{lipsum}

\begin{document}
\title{Spectral Function Method and Janus Quantum Numbers in Quasiperiodic Systems}
  \author{Tian-Le Wu}
  \thanks{These two authors contributed equally.}
	\affiliation{School of Physics and Wuhan National High Magnetic Field Center, Huazhong University of Science and Technology, Wuhan, Hubei 430074, China}

 \author{Shi-Ping Ding}
 \thanks{These two authors contributed equally.}
	\affiliation{School of Physics and Wuhan National High Magnetic Field Center, Huazhong University of Science and Technology, Wuhan, Hubei 430074, China}

  \author{Miao Liang}
  \affiliation{Zhejiang Key Laboratory of Quantum State Control and Optical Field Manipulation, Department of Physics,Zhejiang Sci-Tech University, Hangzhou 310018, China}
  \author{Jing-Tao L{\"u}}    
\email{jtlu@hust.edu.cn}
\affiliation{School of Physics and Wuhan National High Magnetic Field Center, Huazhong University of Science and Technology, Wuhan, Hubei 430074, China}
    \affiliation{Hubei Fundamental Research Center for Physics, Wuhan, Hubei, China.}
 \author{Jin-Hua Gao}
 \email{jinhua@hust.edu.cn}
	\affiliation{School of Physics and Wuhan National High Magnetic Field Center, Huazhong University of Science and Technology, Wuhan, Hubei 430074, China}
    \affiliation{Hubei Fundamental Research Center for Physics, Wuhan, Hubei, China.}
 \begin{abstract}
The absence of translational symmetry in quasiperiodic systems invalidates conventional band theory, posing the central challenge in the field. Building upon the incommensurate energy band (IEB) concept, we establish a unified spectral theory for quasiperiodic systems by introducing two key advances. First, we develop an efficient spectral function method that calculates $A(k,\omega)$ using a small truncated Hamiltonian matrix, bypassing full diagonalization. It converges via a distinctive successive locking of energy moments, yielding exact thermodynamic-limit results without finite-size scaling. Second, we introduce that quasiperiodic eigenstates possess Janus quantum numbers: a single eigenstate, continuously tracked across localization transitions, carries dual labels in momentum and real space, which naturally reduce to the familiar Bloch momentum and band index in the commensurate limit. Together with IEB, these advances constitute a ``band theory'' for quasiperiodic systems, enabling us to define, compute, and label states with the same facility as in periodic ones.
\end{abstract}

\maketitle

\emph{Introduction.}---Quasiperiodic systems, such as the AAH model\cite{harperSingleBandMotion1955,aubry1980}, moir\'e superlattices\cite{bistritzerMoireBandsTwisted2011, caoCorrelatedInsulatorBehaviour2018, caoUnconventionalSuperconductivityMagicangle2018}, have drawn intense research interest over the past few decades. Yet understanding their energy spectra remains a fundamental challenge. The central difficulty is the absence of translational symmetry, which renders Bloch's theorem inapplicable. As a result, the spectrum cannot be described by conventional band theory, even in principle.

Meanwhile, existing numerical methods for quasiperiodic spectra also suffer from significant limitations. Commensurate approximations restore translational symmetry by replacing the quasiperiodic system with an approximately periodic one, enabling a band description, but inherently fail to capture localization. Direct diagonalization in real space requires handling prohibitively large matrices, and the resulting eigenstates carry no momentum-space information, offering limited physical insight.
\begin{figure}[h]
    \centering    
    \includegraphics[width=0.9\linewidth]{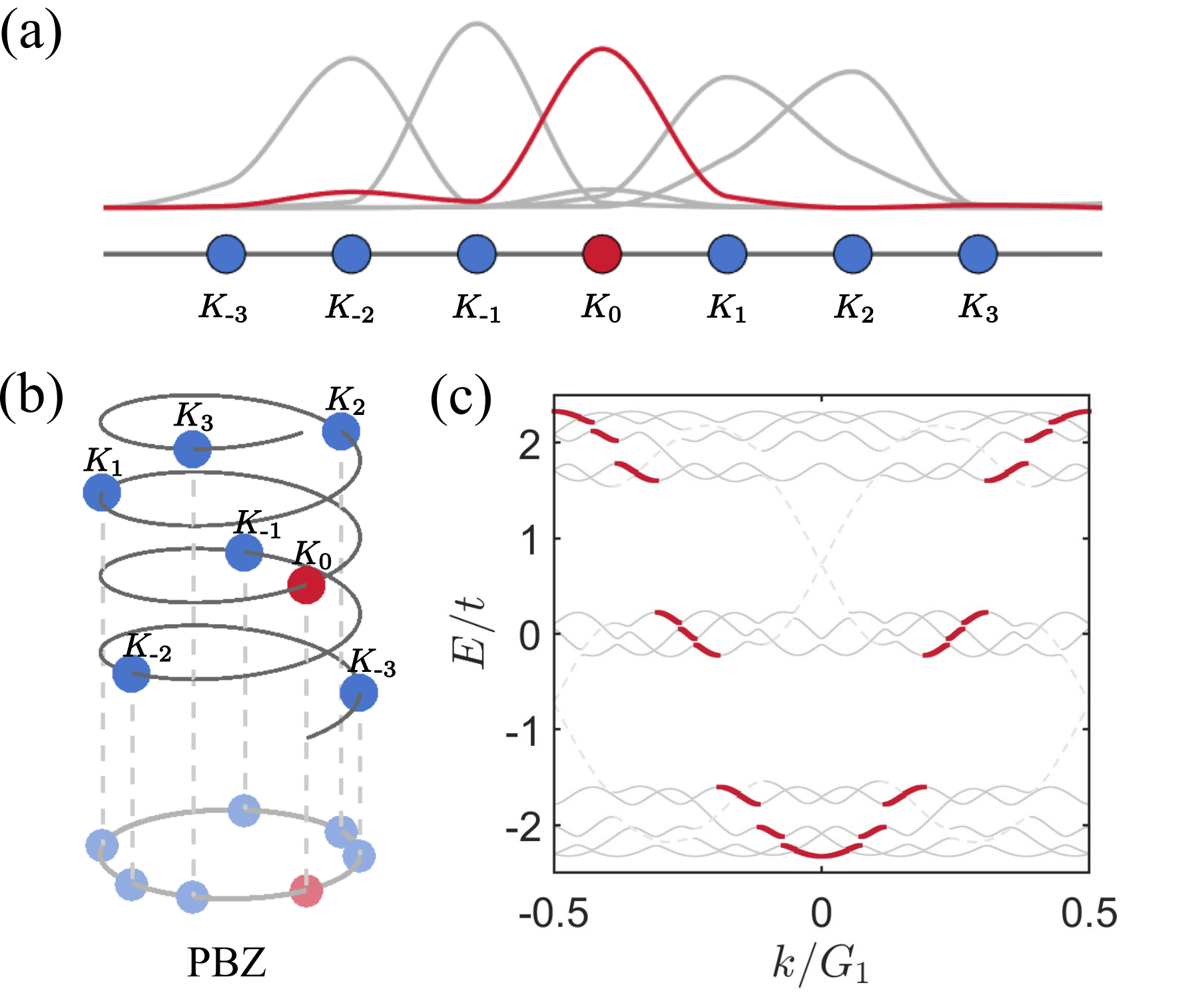}
    \caption{(a) Momentum-space coupling chain of the AAH model. Curves show the wavefunction distribution over momentum sites \(K_m\). (b) Spiral mapping from chain sites \(K_m\) to the PBZ. (c) IEB dispersion (red) and replica bands (gray) at \(V=1.2t\), calculated by the projection method. Throughout this paper, we set \(\alpha=(\sqrt{5}-1)/2\). 
    }
    \label{fig:1}
\end{figure}

Therefore, developing a spectral theory as conceptually clear and computationally simple as band theory is  one of the foremost challenges in quasiperiodic theory\cite{moonQuasicrystallineElectronicStates2019, wangDecomposingElectronicStructures2025, zhouPlaneWaveMethods2019, zhuTwistedTrilayerGraphene2020a}.
Recently, we introduced the concept of incommensurate energy bands (IEB)\cite{heEnergySpectrumTheory2024,guoEnergyBandsIncommensurate2024,chenTheoryLocalizedStates2025}, which generalizes the concept of energy band to quasiperiodic systems and provides a key breakthrough toward resolving this challenge. 
The IEB theory reveals that translational symmetry is not the necessary condition for defining energy bands; rather, it is the localization of eigenstates in the relevant parameter space---here, momentum space. Although quasiperiodic systems lack translational symmetry, their extended states remain localized in momentum space. One can therefore still label the eigenstates by their momentum. This is the concept of IEB\cite{guoEnergyBandsIncommensurate2024}.
However, because eigenstates in complex quasiperiodic systems—though localized—exhibit highly intricate wavefunction distributions in k-space, directly solving the quasiperiodic spectrum via the IEB concept is not universally applicable. Therefore, the calculation of energy spectra for quasiperiodic systems is not yet satisfactorily resolved.

 \begin{figure*}[ht!]
    \centering
    \includegraphics[width=\linewidth]{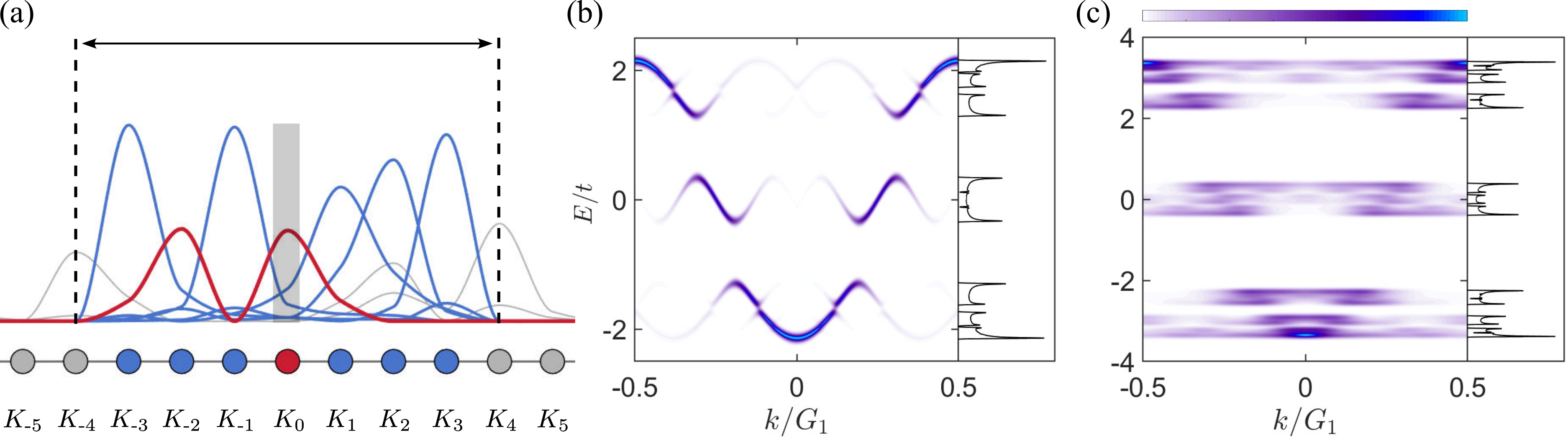}
    \caption{(a) Schematic of the spectral function method: only eigenstates localized near \(|K_0\rangle\) (red, blue) contribute; a double-peak distribution (red) invalidates the projection method. (b) \(A(k,\omega)\) and DOS at \(V=1t\) . (c) \(A(k,\omega)\) and DOS at \(V=3t\). }
    \label{fig:fig2}
\end{figure*}


What is still missing becomes clear when one examines why band theory works so well. Its success rests on three inseparable elements: the concept of energy bands, an efficient computational method, and a complete set of good quantum numbers (crystal momentum). The IEB concept addresses the first. Here we resolve the remaining two.

In this work, we first propose a general and efficient numerical method to calculate the spectral function of quasiperiodic systems. It obtains \(A(k,\omega)\) with controllable accuracy using only a small truncated Hamiltonian matrix, without computing the full set of eigenstates. The method offers four key advantages: (1) \emph{Universality}: generally applicable to quasiperiodic models, covering both extended and localized states, differing only in convergence rate; (2) \emph{Efficiency}: requires diagonalizing only a small truncated matrix, with calculations at different \(k\) points being independent and thus trivially parallelizable; (3) \emph{Rich information}: provides an energy- and momentum-resolved spectral distribution that directly corresponds to ARPES spectra; (4) \emph{Controllable accuracy}: converges numerically through a distinctive successive locking of energy moments, yielding exact values of certain spectral integrals (e.g., average energy) in the thermodynamic limit without finite-size scaling.

Second, we identify a new kind of quantum number in quasiperiodic systems---the Janus quantum number---thereby challenging the long-held view that good quantum numbers cannot be assigned without translational symmetry.
Taking the AAH model as an example, the Wigner--von Neumann theorem\cite{vonneumannUeberMerkwuerdigeDiskrete1993} generically ensures that the energy curves of eigenstates $E_n(V)$ in the $E$-$V$ diagram do not cross as the parameter $V$ varies, allowing each eigenstate to be continuously tracked. When $V<2t$, the eigenstates are localized in momentum space and labeled by a wave vector $k$. When $V > 2t$, they become localized in real space and labeled by a lattice coordinate $r$\cite{chenTheoryLocalizedStates2025}. As $V$ varies, a single eigenstate thus carries two distinct indices---one in momentum space and one in real space---connected by the continuous evolution of the state itself. We term this dual labeling a Janus quantum number. In the commensurate limit, Janus quantum numbers reduce to the familiar Bloch crystal momentum and band index. 

Taken together, the IEB concept introduced in our prior work, and the spectral function method and Janus quantum numbers established here, constitute a complete spectral theory for quasiperiodic systems. This framework makes it possible to define, compute, and label the quantum states of quasiperiodic systems with the same ease that band theory affords for periodic systems.


\emph{A short review of IEB concept.}---We first give a short review for the  concept of IEB using the AAH model, the simplest quasiperiodic system\cite{guoEnergyBandsIncommensurate2024,chenTheoryLocalizedStates2025}. The Hamiltonian of the AAH model is
\begin{equation}\label{AAH_real_space_Lattice}
    H_{\mathrm{AAH}}=t \sum_{j}( c^{\dagger}_{j}c_{j+1}+\mathrm{H.c.})+\sum_jV_jc^{\dagger}_{j}c_{j},
\end{equation}
where $c_j$ is the electron annihilation operator on site $j$ and the quasiperiodic potential $V_j=V \cos(\mathbf{G}_2 \cdot \mathbf{r}_j + \theta )$. Here $\mathbf{G}_2=2\pi/a_2$ and $\mathbf{G}_1=2\pi/a_1$ are the reciprocal lattice vectors of the quasiperiodic potential and the atomic chain, respectively, and $\theta$ is a phase parameter.

A key insight of IEB theory is that extended states are localized in momentum space. We therefore switch to the Bloch basis \(|k\rangle\) of the atomic chain. The quasiperiodic potential \(V_j\) couples Bloch waves separated by \(G_2\), defining the momentum set \(Q_{k}=\{K_m\mid K_m=(k+mG_2)\bmod G_1,\,m\in\mathbb{Z}\}\). Because \(\alpha=G_2/G_1\) is irrational, Kronecker's Theorem\cite{kroneckerNaeherungsweiseGanzzahligeAufloesung1884} and Weyl's Equidistribution Theorem\cite{weylUeberGleichverteilungZahlen1916} guarantee that \(\{K_m\}\) is dense and uniformly distributed in the Brillouin zone (the primary BZ, or PBZ).  Consequently, the momenta throughout the entire BZ can be approximately viewed as all being coupled by  \(V_j\)  into a single momentum-space chain [Fig.~\ref{fig:1}(a)]. 
This chain is described by the tight-binding Hamiltonian
\begin{equation}\label{AAH_k_space_Lattice}
    H(k)=\frac{V}{2} \sum_{m}(c^{\dagger}_{m}c_{m+1}+\mathrm{H.c.})+\sum_{m} T_m c^{\dagger}_{m}c_{m},
\end{equation}
where \(c_m\equiv c_{K_m}\) and \(T_m\equiv 2t\cos(K_m\cdot a_1)\). 
Fig.~\ref{fig:1}(b) illustrates the spiral (modulo) mapping relation between chain sites \(K_m\) and points in the PBZ \cite{chenTheoryLocalizedStates2025}. 
In this picture,  \(k=K_0\)  serves only as the starting point of the momentum chain, so different choices of \(k\)  give the same chain and thus the same physical spectrum.
Equations~\eqref{AAH_real_space_Lattice} and~\eqref{AAH_k_space_Lattice} together manifest the Aubry-Andr\'e self-duality\cite{kohmotoMetalInsulatorTransitionScaling1983, wilkinsonCriticalPropertiesElectron1984}.

The concept of IEB can be understood via an adiabatic picture. From Eq.~\eqref{AAH_k_space_Lattice}, when $V=0$, each eigenstate of $H_{\mathrm{AAH}}$ is simply a Bloch wave $|K_m\rangle$ localized on the $m$-th site of the momentum chain. As $V$ is turned on adiabatically ($V<2t$), each $|K_m\rangle$ evolves into an eigenstate $|\psi_k^{(m)}\rangle$ localized near the $m$-th site, i.e., the eigenstate associated with the momentum $K_m=k+mG_2$ with eigenenergy $E^{(m)}_k$. Each eigenstate can therefore be labeled by the site index $m$ on the momentum chain. Combined with the spiral mapping relation (Fig.~\ref{fig:1}(b)), each eigenstate corresponds uniquely to a point in the PBZ. One can thus define the AAH eigenstates on the PBZ---this is the concept of IEB, and we denote the IEB dispersion by $E(k)$. Since all eigenstates on the momentum chain can be enumerated by either $k$ or $m$, we have the completeness relation\cite{supp}
\begin{equation}\label{completeness relation}
    \sum_{m\in \mathbb{Z}} |\psi_k^{(m)}\rangle \langle \psi_k^{(m)}| = \sum_{k\in \mathrm{PBZ}} |\psi_k^{(0)}\rangle \langle \psi_k^{(0)}|= \mathbf{1}.
\end{equation}

\emph{Projection method for IEB.}---For the AAH model, extended states are sharply localized in momentum space. For example,  the eigenstate associated with \(k\) has its largest amplitude precisely on the basis vector \(|K_0\rangle \equiv |k\rangle\). This suggests a simple computational strategy \cite{chenTheoryLocalizedStates2025, guoEnergyBandsIncommensurate2024}. To obtain the eigenstate for a given \(k\), we construct \(H(k)\) with \(|k\rangle\) as the starting site and truncate to a small window \(\{|K_{-n_c}\rangle,\ldots,|K_0\rangle,\ldots,|K_{n_c}\rangle\}\) that need only cover the extent of the localized wavefunction (red line in Fig.~\ref{fig:1}(a)). Diagonalizing the resulting $N$-dimensional matrix \(H_c(k)\) with  \(N=2n_c+1\), we identify the eigenstate with the largest amplitude on \(|K_0\rangle\) as the target state \(|\psi_k^{(0)}\rangle\), whose eigenvalue is \(E(k)\). Sweeping over all \(k\) in the PBZ yields the complete IEB dispersion $E(k)$, see Fig.~\ref{fig:1}(c). For \(V>2t\), by duality the same procedure applied to the real-space lattice yields the \emph{localized-state energy band} (LSEB) \(E(r)\) in the real-space Brillouin zone (RBZ)\cite{chenTheoryLocalizedStates2025}.

The projection method assumes a single-peak structure: each eigenstate must have a dominant amplitude on its associated basis vector. While this holds for nearly all eigenstates in the AAH model, it fails in more complex quasiperiodic systems such as twisted bilayer graphene (TBG), where intricate couplings spread an eigenstate over multiple basis vectors. For example, near degeneracy points, the gap-opening hybridizes the two basis vectors, producing nearly equal amplitudes on both (red line in Fig.~\ref{fig:fig2}(a)), which invalidates the ``largest amplitude'' labeling criterion used to determine the complete IEB dispersion. The projection method is therefore not universally applicable.

\emph{Spectral function method.}---
Computing the spectral function is a better approach. Still taking the AAH model
as an example, the spectral function in the Bloch representation is defined as
\begin{equation}
A(k,\omega) = \langle k | \delta(\omega - H_{\mathrm{AAH}}) | k \rangle,
\end{equation}
where \(|k\rangle\) is the chosen Bloch basis vector, and $\omega$ is the single-particle energy in units $\hbar=1$. Here, without loss of
generality, we set \(k\) as the starting point of the momentum chain,
i.e., \(|k\rangle \equiv |K_0\rangle\). Inserting the completeness relation immediately yields
\begin{equation}
\begin{aligned}
    A(k,\omega) &= \sum_m |\langle K_0 | \psi^{(m)}_k\rangle|^2 \, \delta(\omega - E^{(m)}_k) \\
     &= \sum_m |C^{(m)}_0(k)|^2 \, \delta(\omega - E^{(m)}_{k}),
\end{aligned}
\label{Afunction}
\end{equation}
where \(C^{(m)}_{n}(k)\) is the coefficient of \(|\psi^{(m)}_k\rangle\) on \(|K_n\rangle\) with
\begin{equation}
    \ket{\psi^{(m)}_{k}} = \sum_n C^{(m)}_{n}(k)\ket{K_n}.
\end{equation}
Eq.~\eqref{Afunction} is the central formula of spectral function method. 

The great advantage of Eq.~\eqref{Afunction} is that the localization  of eigenstates actually allows us to compute \(A(k,\omega)\) accurately with a very small matrix truncation. As shown in Fig.~\ref{fig:fig2}(a), the eigenstates \(|\psi^{(m)}_k\rangle\) that contribute to \(|K_0\rangle\) are just the  ones localized near \(|K_0\rangle\) (blue and red lines), while  states localized far from \(|K_0\rangle\) (gray lines) contribute negligibly. Therefore, as long as the truncation \(n_c\) encompasses these contributing states, \(A(k,\omega)\) is obtained accurately. Importantly, unlike the projection method, Eq.~\eqref{Afunction} is free of the labeling ambiguity of eigenstates, since it sums over all contributing eigenstates with their natural weights.

\begin{figure*}
    \centering
    \includegraphics[width=0.95\linewidth]{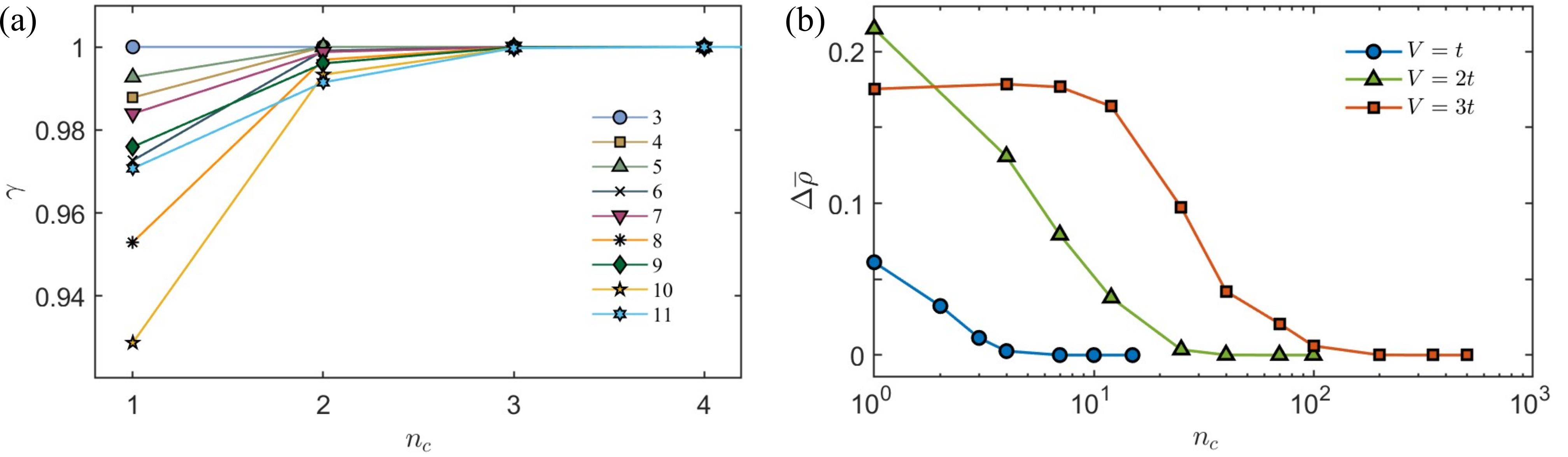}
    \caption{(a) Moment-locking ratio \(\gamma=\mu_{p,c}(k)/\mu_p(k)\) vs.\ \(n_c\) for \(p=3\)--\(11\), \(V=1t\), \(k=0.1G_1\). Lower-order moments lock to unity at small \(n_c\); higher orders require progressively larger truncations\cite{supp}. (b) Density of states convergence error \(\Delta \overline{\rho}\)\cite{supp} vs.\ \(n_c\) (log scale) for \(V=t\) (extended), \(V=2t\) (critical), and \(V=3t\) (localized).}
    \label{fig:fig3}
\end{figure*}

The computational procedure is thus as follows. Take a truncation $n_c$ to form the truncated matrix \(H_c(k)\). Diagonalize \(H_c(k)\) to obtain all eigenstates \(|\psi^{(m)}_k\rangle\). Compute \(A(k,\omega)\) via Eq.~\eqref{Afunction}. Finally, sweep over all \(k \in \mathrm{PBZ}\) to obtain the complete spectral function, and check the convergence of \(n_c\) using the density of states\cite{supp}
\begin{equation}
    \rho(\omega) =  \int_{\mathrm{PBZ}} \frac{dk}{2\pi}A(k,\omega) .
\end{equation}
Fig.~\ref{fig:fig2} (b)  plot the $A(k,\omega)$ of AAH model with $V=1t$, which agrees exactly with the projection method. Interestingly, even for $V>2t$, the Eq.~\eqref{Afunction} is also valid as long as the trunction $n_c$ is large enough, see Fig.~\ref{fig:fig2}(c) . 

\emph{Convergence analysis.}---Let \(A_c(k,\omega)\) be the spectral function obtained from the \(N\)-dimensional truncated matrix \(H_c(k)\) via Eq.~\eqref{Afunction}. As we will show, with increased truncation, \(A_c(k,\omega)\) approaches the exact thermodynamic-limit solution through a distinctive successive locking of energy moments from low to high order: low-order moments determine the coarse-grained structure of \(A(k,\omega)\) and dominate, while high-order moments contribute only finer, rapidly decaying details. The truncation $n_c$ therefore sets the energy resolution of the spectral function.

To make this precise, consider the characteristic function
\begin{equation}
    \chi(t) = \int e^{it\omega} A(k,\omega) \, d\omega 
            = \langle K_0 | e^{itH} | K_0 \rangle
            = \sum_{p=0}^{\infty} \frac{(it)^p}{p!} \, \mu_p,
\end{equation}
where the energy moments are
\begin{equation}\label{eq:moments}
    \mu_p(k) \equiv \int \omega^{p} A(k,\omega)\,d\omega = \bra{K_0} H^{p}(k) \ket{K_0}.
\end{equation}
Inverse Fourier transform of \(\chi(t)\) gives the moment expansion\cite{supp}
\begin{equation}
\begin{aligned}
    A(k,\omega) &= \sum_{p=0}^{\infty} \frac{(-1)^{p}}{p!} \, \mu_{p}(k) \, \delta^{(p)}(\omega) \\[4pt]
                &= G_{\sigma}(\omega)\,
                   \sum_{p=0}^{\infty} 
                   \frac{\mu_{p}(k)}{p!\,\sigma^{p}}\,
                   \mathrm{He}_{p}\!\left(\frac{\omega}{\sigma}\right),
\end{aligned}
\label{eq:spectral-Gauss}
\end{equation}
where \(\delta^{(p)}(\omega)\) is the \(p\)-th derivative of the \(\delta\)-function, and in the second line we replaced \(\delta(\omega)\) by a Gaussian \(G_{\sigma}(\omega)=\exp(-\omega^2/2\sigma^2)/\sqrt{2\pi}\sigma\), with \(\mathrm{He}_{p}(x)=(-1)^p e^{x^2/2}\frac{d^p}{dx^p}e^{-x^2/2}\) the probabilists' Hermite polynomials. Thus, the spectral function is uniquely determined by the full set of energy moments \(\{\mu_p\}\).

A key fact is that a finite truncated $H_c$ yields exact low-order moments. It follows from a simple tightbinding (TB) picture\cite{cyrot-lackmannElectronicStructureLiquid1967,DUCASTELLE19701295}. A TB Hamiltonian \(H\) contains only nearest-neighbor hopping, so each application of \(H\) moves the electron by at most one step (here is in $k$ space). \(\mu_p=\langle K_0|H^p|K_0\rangle\) is the sum of amplitudes over all \(p\)-step 
closed paths starting and ending at \(|K_0\rangle\)---it involves only sites within \(p\) steps of \(|K_0\rangle\). As long as the truncation radius \(N\ge p\), the electron never encounters the boundary. Hence, \(H_c\) reproduces the first \(N\) moments $\mu_p$ exactly. Enlarging \(N\) locks in ever higher-order moments, and 
as \(N\to\infty\) all moments are locked, guaranteeing convergence of \(A_c(k,\omega)\).  Fig.~\ref{fig:fig3} (a) shows the calculated \(\mu_p\) versus \(n_c\): low-order moments converge at very small \(n_c\), while high-order moments require larger \(n_c\)---a direct visualization of successive locking.


\begin{figure*}
    \centering
    \includegraphics[width=0.95\linewidth]{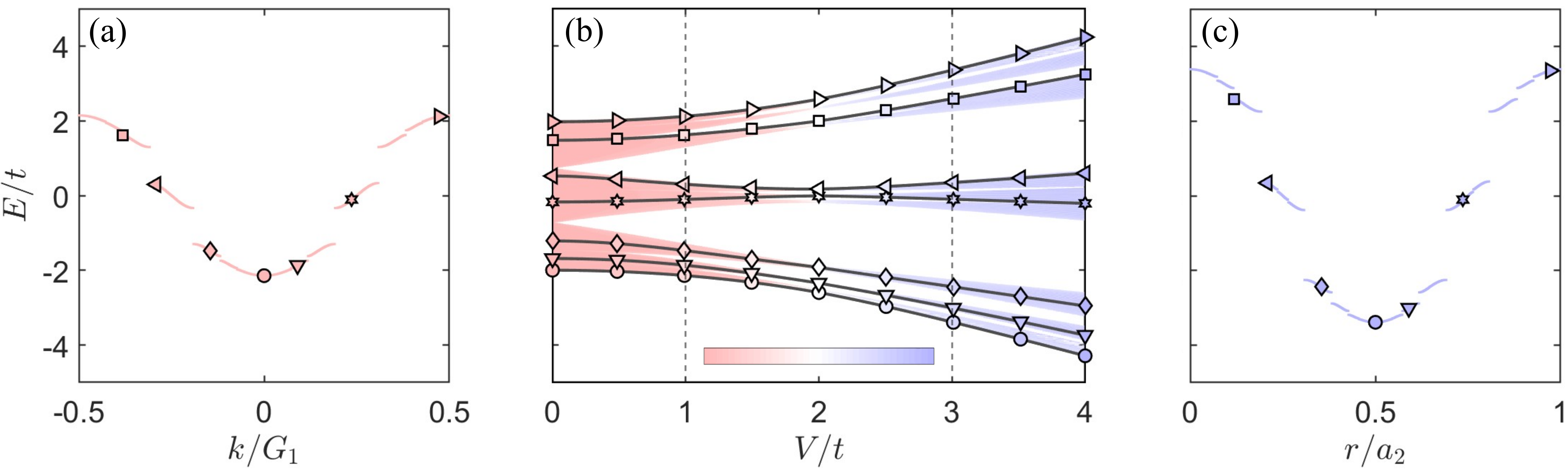}
    \caption{Janus quantum numbers in the AAH model. (a) IEB dispersion at \(V=1t\): seven eigenstates (\(E\) vs.\ \(k\)) marked by distinct symbols. (b)  IPR diagram of AAH model\cite{supp}. Seven non-crossing curves track the same eigenstates continuously from \(V<2t\) to \(V>2t\). (c) LSEB at \(V=3t\)\cite{supp}: the same seven eigenstates (\(E\) vs.\ \(r\)) now labeled by real-space coordinate \(r\), revealing the dual face of a Janus quantum number.}
    \label{fig:janus}
\end{figure*}

Eq.~\eqref{eq:spectral-Gauss} reveals that high-order moments encode finer energy resolution. Note that \(\mathrm{He}_p(\omega/\sigma)\) has \(p\) nodes in \([-\sqrt{2p},\sqrt{2p}]\), resolving features on the scale \(\sigma/\sqrt{p}\). Meanwhile, the coefficient \(\mu_p/(p!\,\sigma^p)\) is rapidly suppressed by the  factorial \(p!\). Therefore,  truncation sacrifices only fine high-resolution details, which are naturally small\cite{supp}. In other words, low-order moments $\mu_p$ determine the coarse-grained structure and dominate, so that, for a given energy resolution, a finite truncation suffices.

Note that Eq.~\eqref{Afunction} and Eq.~\eqref{eq:spectral-Gauss} are two equivalent representations of \(A(k,\omega)\) derived from the same \(H_c(k)\):  the former is the most convenient for numerical computation; the latter exposes the successive-locking structure and provides a clear convergence criterion. For any given 
truncation \(N\), the two formulas yield identical \(A_c(k,\omega)\). Crucially, Eq.~\eqref{eq:spectral-Gauss} makes no assumption about whether the underlying eigenstates are extended or localized; it implies that Eq.~\eqref{Afunction} applies equally to localized states, provided the truncation of \(H_c\) is sufficiently large. As shown in Fig.~\ref{fig:fig3}(b), extended states (\(V<2t\)) require very small \(n_c\) because their eigenstates are localized in momentum space and the contribution of high-order moments is negligible; critical states (\(V=2t\)) and localized states (\(V>2t\)) require \(n_c\) one to two orders of magnitude larger to converge.

Moment locking also yields a distinctive advantage 
without finite-size scaling: if \(H_c\) locks the first \(M\) moments, then for any 
polynomial \(Q(\omega)\) of degree at most \(M\),
\begin{equation}\label{eq:intQ}
    \int Q(\omega)\,A_c(k,\omega)\,d\omega \;=\; \int Q(\omega)\,A(k, \omega)\,d\omega
\end{equation}
holds exactly\cite{supp}. Average energy and energy fluctuations can thus be computed rigorously from the truncated matrix, without any need of  finite-size scaling.  And even low-temperature heat 
capacity can be obtained very accurately in the same way. 

\emph{Janus quantum numbers.}---The IEB concept assigns a momentum label \(k\) to each eigenstate when \(V<2t\); the LSEB concept \cite{chenTheoryLocalizedStates2025} assigns a real-space label \(r\) when \(V>2t\). We now show that 
\(k\) and \(r\) are not two independent quantum numbers (labels of the eigenstates), but two faces of the same eigenstate's identity in different parameter regimes. This identification rests on the Wigner--von Neumann non-crossing theorem: for a real symmetric matrix \(H(V)\) depending on a single parameter \(V\), the eigenvalues \(E_n(V)\) generically do not cross \cite{vonneumannUeberMerkwuerdigeDiskrete1993}. Each eigenstate can therefore be continuously tracked as \(V\) varies. Take an 
eigenstate at \(V<2t\), labeled by its momentum-space index \(k\); increase \(V\)  across the localization transition at \(V=2t\) into the localized phase. The same eigenstate, now localized in real space, is labeled by a lattice coordinate \(r\). The two labels are thus placed in one-to-one correspondence through the eigenstate itself: \(k \leftrightarrow r\)\cite{supp}. We term this dual-labeling structure 
a \emph{Janus quantum number}.

Fig.~\ref{fig:janus} illustrates this explicitly for the AAH model. 
Fig.~\ref{fig:janus}(a) shows the IEB at \(V=1t\) with several eigenstates 
marked by their momenta. Fig.~\ref{fig:janus}(b) tracks these same eigenstates 
in the \(E\)-\(V\) diagram, color-coded by the inverse participation ratio (IPR); 
the non-crossing curves are clearly visible, and the IPR shows the continuous 
crossover from extended to localized behavior at \(V=2t\). Fig.~\ref{fig:janus}(c) 
plots the LSEB at \(V=3t\), where the same eigenstates are now naturally labeled 
by real-space indices---a direct visualization of the Janus property.

It is important to note that the two indices are not simultaneously sharp.  When \(V<2t\), \(k\) is a sharp label (the eigenstate is localized in momentum space) while \(r\) is a blurred label (the eigenstate is extended in  real space); when \(V>2t\), the roles reverse. One face is sharp, the other blurred---yet they transform into each other through the continuous variation of \(V\) and correspond one-to-one. We conjecture that this structure generalizes beyond the AAH model: for any quasiperiodic Hamiltonian depending on a single parameter and satisfying the Wigner–von Neumann condition, Janus quantum numbers can be defined.


In the commensurate limit, where the modulation becomes commensurate and translational symmetry is restored, eigenstates are simultaneously extended in real space and \(\delta\)-function localized in momentum space. The two faces of the Janus quantum number then collapse into one---the familiar Bloch crystal momentum 
\(k\) and band index \(n\). Bloch quantum numbers are therefore a degenerate special case of Janus quantum numbers.

Janus quantum numbers reveal a deeper principle: the physical origin of quantum numbers is localization in parameter space, not symmetry. Fourier duality guarantees that every eigenstate exists simultaneously in conjugate spaces;  whichever space it happens to be localized in provides the natural coordinate to label it. Symmetry is merely one extreme mechanism that guarantees such localization---as in Bloch's theorem, where translational symmetry forces each 
eigenstate onto a single \(k\)-point. In quasiperiodic systems, where symmetry  is absent, localization alone suffices to define good quantum numbers.

\emph{Conclusion.}---
In summary, we have established a unified spectral theory for quasiperiodic systems, combining the IEB concept, an efficient spectral function method based on moment locking, and Janus quantum numbers. Together, these advances make it possible to define, compute, and label quasiperiodic states with the same facility that Bloch theory affords for periodic systems.
 

\begin{acknowledgments}
    This work was supported by the National Natural Science Foundation of China (Grants No.~12141401 and No.~22273029), the National Key Research and Development Program of China (Grants No.~2022YFA1403501 and No.~2022YFA1402400), China Postdoctoral Science Foundation (Grant No. 2024M750984), and Innovation Program for Quantum Science and Technology (Grant No. 2021ZD0302400). 
\end{acknowledgments}

\bibliography{reference}

\clearpage
\onecolumngrid

\clearpage
\onecolumngrid

\setcounter{equation}{0}
\setcounter{figure}{0}
\setcounter{table}{0}
\setcounter{page}{1}
\makeatletter
\renewcommand{\theequation}{S\arabic{equation}}
\renewcommand{\thefigure}{S\arabic{figure}}
\renewcommand{\thetable}{S\arabic{table}}
\renewcommand{\bibnumfmt}[1]{[S#1]}

\begin{center}
\textbf{\large Supplementary Material for Spectral Function Method and Janus Quantum Numbers in Quasiperiodic Systems}
\end{center}

\section{I. Proof of the Completeness Relation}

According to the spiral mapping (Fig.~\ref{fig:1}(b)), the discrete $k$-space lattice sites are in one-to-one correspondence with continuous momenta in the PBZ. Suppose we choose another momentum $k'=(k+nG_2)\bmod G_1$ in the PBZ and write down the Hamiltonian $H(k')$. In Fig.~1(a), this corresponds to shifting the central site (red sphere) from $K_0$ to $K_n$. For this one-dimensional infinite $k$-space lattice, the choice of the central site does not alter any physical essence:
\begin{equation}
    H=H(k)=H(k'),\quad k,k'\in \mathrm{PBZ}.
    \tag{S1}
\end{equation}
We thus conclude: solving the eigenvalue equation of $H(k)$ at any chosen momentum $k$ yields the complete energy spectrum of the AAH model, and the resulting set of eigenstates satisfies completeness:
\begin{equation}\label{eq:single_k_completeness}
    \sum_{m\in\mathbb{Z}} \ket{\psi^{(m)}_k}\bra{\psi^{(m)}_k} = \mathbf{1}.
    \tag{S2}
\end{equation}

Eigenstates at different momenta $k$ are related by the energy-band translation relation:
\begin{equation}\label{eq:translation}
    \ket{\psi^{(n+m)}_{k}} = \ket{\psi^{(m)}_{k+nG_2}}.
    \tag{S3}
\end{equation}
This relation can be rigorously understood from the adiabatic definition of the eigenstates. By construction, the eigenstate $\ket{\psi^{(m)}_k}$ is obtained as the final state of an adiabatic evolution starting from the initial state $\ket{k+mG_2}$ under the Hamiltonian $H(k)$ corresponding to the momentum parameter $k$. Hence, the initial state of $\ket{\psi^{(n+m)}_k}$ is $\ket{k+(n+m)G_2}$, while the initial state of $\ket{\psi^{(m)}_{k+nG_2}}$ is $\ket{(k+nG_2)+mG_2}=\ket{k+(n+m)G_2}$. Clearly, both originate from exactly the same initial Bloch state $\ket{K_{n+m}}$. Moreover, since $H(k)=H(k+nG_2)$, the Hamiltonians along the two adiabatic paths are identical. The adiabatic theorem guarantees that the final state is uniquely determined by the initial state and the evolution path; therefore, these two processes necessarily produce the same final state, rigorously establishing the energy-band translation relation.

Setting $m=0$ in Eq.~\ref{eq:translation} and relabeling $n$ as $m$, we obtain
\begin{equation}
    \ket{\psi^{(m)}_{k}} = \ket{\psi^{(0)}_{K_m}}.
    \tag{S4}
\label{eq:one_to_one}
\end{equation}
This equation establishes a one-to-one correspondence between two sets of eigenstates: the left-hand side fixes the momentum $k$ and traverses the band index $m$, while the right-hand side fixes the band index $0$ and traverses all discrete momenta $K_m = k+mG_2$. Substituting Eq.~\ref{eq:one_to_one} into Eq.~\ref{eq:single_k_completeness} yields
\begin{equation}
    \sum_{m\in\mathbb{Z}} \ket{\psi^{(0)}_{K_m}}\bra{\psi^{(0)}_{K_m}} = \mathbf{1}.
    \tag{S5}
\end{equation}
Since $\{K_m\}$ is dense in the PBZ, the above discrete sum is equivalent to a continuous momentum integral. Relabeling the discrete momentum $K_m$ as the continuous momentum $k\in\mathrm{PBZ}$, we obtain the completeness relation for the incommensurate energy band:
\begin{equation}\label{eq:completeness}
    \sum_{m\in\mathbb{Z}} \ket{\psi^{(m)}_k}\bra{\psi^{(m)}_k} = \sum_{k\in\mathrm{PBZ}} \ket{\psi^{(0)}_k}\bra{\psi^{(0)}_k} = \mathbf{1}.
    \tag{S6}
\end{equation}

We define the dispersion relation $E(k)=E^{(0)}_k$ as the incommensurate energy band (IEB). The choice of the band index $m=0$ as the IEB carries special significance, because $\ket{\psi^{(0)}_k}$ evolves adiabatically from the reference state $\ket{K_0=k}$ and corresponds to the genuine physical energy band. The remaining bands $E^{(m\neq 0)}_k$ are called replica bands; they are redundant representations arising from the $G_2$ periodicity of $H(k)$.

Eq.~\ref{eq:completeness} shows that there are two equivalent viewpoints for describing the band structure of incommensurate systems: (i)~fix a momentum $k$ and traverse all band indices $m$; (ii)~fix the band index $m=0$ and traverse all momenta $k$ in the PBZ. These correspond precisely to the reduced-zone scheme and the extended-zone scheme in conventional band theory. Within this framework, incommensurate and commensurate energy bands can be unified in the extended-zone picture: both appear as a single band in the PBZ. The band index $m$ in a commensurate system can be relabeled analogously, taking values in a finite periodic sequence $m=0,1,\dots,q-1$, where $q$ is the denominator of the commensurate ratio.

\section{II. Derivation of the Density of States Formula}

According to Weyl's equidistribution theorem\cite{weylUeberGleichverteilungZahlen1916} and Birkhoff's ergodic theorem\cite{birkhoffProofErgodicTheorem1931}, the following relation holds:
\begin{equation}
    \lim_{n_c\rightarrow \infty}\frac{1}{N}\sum_{n=-n_c}^{n_c} f(k+nG_2)=\frac{1}{G_1}\int_{\mathrm{PBZ}} f(k) \, dk,
    \tag{S7}
\end{equation}
where $f$ is a function with period $G_1$, and $N=2n_c+1$. Accordingly, the density of states of an incommensurate system can be written as
\begin{equation}
    \rho(\omega)=\frac{1}{L}\sum_{k\in \mathrm{PBZ}}A(k,\omega)
    =\frac{N_k}{LG_1}\int_{\mathrm{PBZ}}dk\, A(k,\omega)=\int_{\mathrm{PBZ}}\frac{dk}{2\pi}\, A(k,\omega),
    \tag{S8}
\end{equation}
where $L=N_k a_1$, $N_k$ is the number of $k$-points sampled in the PBZ, and $a_1$ is the real-space unit-cell length. The above expression is formally identical to the result for a commensurate system, yet its derivation does not rely on periodic boundary conditions.

When the projection method can successfully locate the IEB, there is a simpler way to obtain the density of states:
\begin{equation}
    \rho(\omega)=\frac{1}{L}\sum_{k\in\mathrm{PBZ}}\delta(\omega-E^{(0)}_k).
    \tag{S9}
\end{equation}

\section{III. Numerical Demonstration of the Moment-Locking}
We define the moment locking ratio
$\gamma_p(n_c) \equiv \mu_{p,c}(k)/\mu_p(k)$ to quantify how closely
the truncated moment approaches the true one. Here
$\mu_{p,c}(k)$ is the $p$-th moment evaluated from the truncated
Hamiltonian $H_c(k)$, and $\mu_p(k)$ is the true moment, obtained
from a sufficiently large truncation once convergence has been
confirmed.  Via Eq.~(5), $\mu_{p,c}(k)$ can be computed in two
equivalent ways:
\begin{equation}
    \mu_{p,c}(k)=\sum_{m\in[-n_c,n_c]}|C^{(m)}_0(k)|^2\bigl(E^{(m)}_k\bigr)^p
    =\langle K_0|H_c^{\,p}(k)|K_0\rangle.
    \tag{S10}
\end{equation}
Numerical tests show that the relative difference between the two
evaluations is negligible (below $10^{-10}$).  The latter, involving only matrix operations on $H_c$, is computationally cheaper and is
therefore adopted throughout this work.

\begin{figure}[H]
    \centering
    \includegraphics[width=0.6\linewidth]{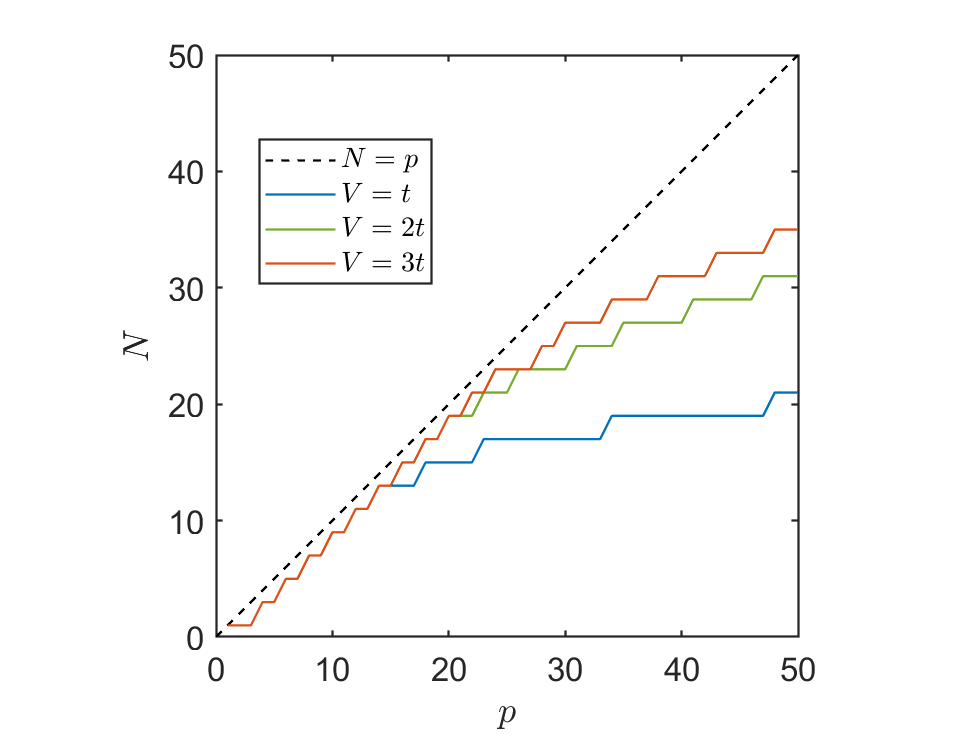}
    \caption{Moment-locking threshold curve.}
    \label{fig:lock}
\end{figure}

As the matrix truncation increases, the moments converge rapidly to extremely high precision. Tab.~\ref{tab:tab1} shows the data of Fig.~\ref{fig:fig3}(a). Red regions indicate moments not yet locked; green regions indicate locked moments. Taking a relative error below $10^{-10}$ as the threshold, we record the matrix truncation at which each moment locks and plot Fig.~\ref{fig:lock}. The black dashed line marks the theoretical bound $N = p$\cite{gautschiConstructionGaussChristFei}; the colored curves show the actual locking thresholds for different values of $V$.  All simulated curves lie on or below the theoretical line, indicating that moment locking occurs earlier than the theoretical bound predicts---consistent with theory.

\begin{sidewaystable}
    \centering
\caption{Moments $\mu_p$ at various truncations $n_c$ for $V=1t$ and $k=0.1G_2$.}
\label{tab:tab1}
\begin{tabular}{|c|*{5}{Wc{4cm}|}}
\hline
\diagbox[width=2.5em, height=2.5em]{$n_c$}{$p$}  & $3$ & $4$ & $5$ & $6$ & $7$  \\ \hline
1 & \textcolor{green!60!black}{\num{-5.2575580128e0}} & \textcolor{red!80!black}{\num{1.0127712199e1}} & \textcolor{red!80!black}{\num{-1.6839814292e1}} & \textcolor{red!80!black}{\num{3.3081963924e1}} & \textcolor{red!80!black}{\num{-5.3707916309e1}}  \\ \hline
2 & \textcolor{green!60!black}{\num{-5.2575580128e0}} & \textcolor{green!60!black}{\num{1.0252712199e1}} & \textcolor{green!60!black}{\num{-1.6963733037e1}} & \textcolor{red!80!black}{\num{3.3981482418e1}} & \textcolor{red!80!black}{\num{-5.4519649210e1}}  \\ \hline
3 & \textcolor{green!60!black}{\num{-5.2575580128e0}} & \textcolor{green!60!black}{\num{1.0252712199e1}} & \textcolor{green!60!black}{\num{-1.6963733037e1}} & \textcolor{green!60!black}{\num{3.4012732418e1}} & \textcolor{green!60!black}{\num{-5.4585814288e1}}  \\ \hline
4 & \textcolor{green!60!black}{\num{-5.2575580128e0}} & \textcolor{green!60!black}{\num{1.0252712199e1}} & \textcolor{green!60!black}{\num{-1.6963733037e1}} & \textcolor{green!60!black}{\num{3.4012732418e1}} & \textcolor{green!60!black}{\num{-5.4585814288e1}}  \\ \hline
5 & \textcolor{green!60!black}{\num{-5.2575580128e0}} & \textcolor{green!60!black}{\num{1.0252712199e1}} & \textcolor{green!60!black}{\num{-1.6963733037e1}} & \textcolor{green!60!black}{\num{3.4012732418e1}} & \textcolor{green!60!black}{\num{-5.4585814288e1}}  \\ \hline
6 & \textcolor{green!60!black}{\num{-5.2575580128e0}} & \textcolor{green!60!black}{\num{1.0252712199e1}} & \textcolor{green!60!black}{\num{-1.6963733037e1}} & \textcolor{green!60!black}{\num{3.4012732418e1}} & \textcolor{green!60!black}{\num{-5.4585814288e1}}  \\ \hline
7 & \textcolor{green!60!black}{\num{-5.2575580128e0}} & \textcolor{green!60!black}{\num{1.0252712199e1}} & \textcolor{green!60!black}{\num{-1.6963733037e1}} & \textcolor{green!60!black}{\num{3.4012732418e1}} & \textcolor{green!60!black}{\num{-5.4585814288e1}}  \\ \hline
8 & \textcolor{green!60!black}{\num{-5.2575580128e0}} & \textcolor{green!60!black}{\num{1.0252712199e1}} & \textcolor{green!60!black}{\num{-1.6963733037e1}} & \textcolor{green!60!black}{\num{3.4012732418e1}} & \textcolor{green!60!black}{\num{-5.4585814288e1}}  \\ \hline
9 & \textcolor{green!60!black}{\num{-5.2575580128e0}} & \textcolor{green!60!black}{\num{1.0252712199e1}} & \textcolor{green!60!black}{\num{-1.6963733037e1}} & \textcolor{green!60!black}{\num{3.4012732418e1}} & \textcolor{green!60!black}{\num{-5.4585814288e1}}  \\ \hline
10 & \textcolor{green!60!black}{\num{-5.2575580128e0}} & \textcolor{green!60!black}{\num{1.0252712199e1}} & \textcolor{green!60!black}{\num{-1.6963733037e1}} & \textcolor{green!60!black}{\num{3.4012732418e1}} & \textcolor{green!60!black}{\num{-5.4585814288e1}}  \\ \hline
\end{tabular}

\bigskip\bigskip

\centering
\begin{tabular}{|c|*{5}{Wc{4cm}|}}
\hline
\diagbox[width=2.5em, height=2.5em]{$n_c$}{$p$}  & $8$ & $9$ & $10$ & $11$ & $12$  \\ \hline
1 & \textcolor{red!80!black}{\num{1.0843183073e2}} & \textcolor{red!80!black}{\num{-1.7051654468e2}} & \textcolor{red!80!black}{\num{3.5684878835e2}} & \textcolor{red!80!black}{\num{-5.3809058767e2}} & \textcolor{red!80!black}{\num{1.1803955765e3}}  \\ \hline
2 & \textcolor{red!80!black}{\num{1.1343271597e2}} & \textcolor{red!80!black}{\num{-1.7403047391e2}} & \textcolor{red!80!black}{\num{3.8167047280e2}} & \textcolor{red!80!black}{\num{-5.4958413451e2}} & \textcolor{red!80!black}{\num{1.2965895159e3}}  \\ \hline
3 & \textcolor{red!80!black}{\num{1.1377617581e2}} & \textcolor{red!80!black}{\num{-1.7471188454e2}} & \textcolor{red!80!black}{\num{3.8413264512e2}} & \textcolor{red!80!black}{\num{-5.5415905601e2}} & \textcolor{red!80!black}{\num{1.3110547976e3}}  \\ \hline
4 & \textcolor{green!60!black}{\num{1.1378398831e2}} & \textcolor{green!60!black}{\num{-1.7472366936e2}} & \textcolor{red!80!black}{\num{3.8423414882e2}} & \textcolor{red!80!black}{\num{-5.5431701698e2}} & \textcolor{red!80!black}{\num{1.3119104932e3}}  \\ \hline
5 & \textcolor{green!60!black}{\num{1.1378398831e2}} & \textcolor{green!60!black}{\num{-1.7472366936e2}} & \textcolor{green!60!black}{\num{3.8423610194e2}} & \textcolor{green!60!black}{\num{-5.5431951772e2}} & \textcolor{red!80!black}{\num{1.3119401548e3}}  \\ \hline
6 & \textcolor{green!60!black}{\num{1.1378398831e2}} & \textcolor{green!60!black}{\num{-1.7472366936e2}} & \textcolor{green!60!black}{\num{3.8423610194e2}} & \textcolor{green!60!black}{\num{-5.5431951772e2}} & \textcolor{green!60!black}{\num{1.3119406431e3}}  \\ \hline
7 & \textcolor{green!60!black}{\num{1.1378398831e2}} & \textcolor{green!60!black}{\num{-1.7472366936e2}} & \textcolor{green!60!black}{\num{3.8423610194e2}} & \textcolor{green!60!black}{\num{-5.5431951772e2}} & \textcolor{green!60!black}{\num{1.3119406431e3}}  \\ \hline
8 & \textcolor{green!60!black}{\num{1.1378398831e2}} & \textcolor{green!60!black}{\num{-1.7472366936e2}} & \textcolor{green!60!black}{\num{3.8423610194e2}} & \textcolor{green!60!black}{\num{-5.5431951772e2}} & \textcolor{green!60!black}{\num{1.3119406431e3}}  \\ \hline
9 & \textcolor{green!60!black}{\num{1.1378398831e2}} & \textcolor{green!60!black}{\num{-1.7472366936e2}} & \textcolor{green!60!black}{\num{3.8423610194e2}} & \textcolor{green!60!black}{\num{-5.5431951772e2}} & \textcolor{green!60!black}{\num{1.3119406431e3}}  \\ \hline
10 & \textcolor{green!60!black}{\num{1.1378398831e2}} & \textcolor{green!60!black}{\num{-1.7472366936e2}} & \textcolor{green!60!black}{\num{3.8423610194e2}} & \textcolor{green!60!black}{\num{-5.5431951772e2}} & \textcolor{green!60!black}{\num{1.3119406431e3}}  \\ \hline
\end{tabular}

\end{sidewaystable}

\section{IV. Definition of the Absolute Error of the Density of States}

We define the absolute error of the density of states as
\begin{equation}
    \Delta\overline{\rho}=\frac{1}{N_E}\sum_E|\rho_c(E)-\rho(E)|,
    \tag{S11}
\end{equation}
where $N_E$ is the number of uniformly spaced energy points, $\rho_c(E)$ is the density of states computed from the finite truncated Hamiltonian matrix, and $\rho(E)$ is the true density of states, approximated by the result obtained with the largest truncation.

\section{V. Derivation of the Spectral Function in Moment Representation}

The spectral function $A(k,\omega)$ can be expressed as the inverse Fourier transform of the characteristic function $\chi(t)$:
\begin{equation}\label{eq:pu_ju}
\begin{aligned}
    A(k,\omega)&=\frac{1}{2\pi}\int e^{-it\omega}\chi(t)dt\\
    &=\frac{1}{2\pi}\int e^{-it\omega} \sum_{p=0}^{\infty}\frac{\mu_p(k)}{p!}(it)^pdt\\
    &=\sum_{p=0}^{\infty} \frac{(-1)^{p}}{p!}\mu_p(k)\frac{1}{2\pi}\int (-it)^pe^{-it\omega}dt\\
    &=\sum_{p=0}^{\infty} \frac{(-1)^{p}}{p!} \, \mu_{p}(k) \, \delta^{(p)}(\omega),
\end{aligned}
\tag{S12}
\end{equation}
where $\delta^{(p)}(\omega)$ is the $p$-th derivative of $\delta(\omega)$:
\begin{equation}
    \delta^{(p)}(\omega)=\frac{1}{2\pi}\int (-it)^pe^{-it\omega}dt.
    \tag{S13}
\end{equation}

Replacing the $\delta$-function by a Gaussian $G_{\sigma}(\omega)=\frac{1}{\sqrt{2\pi}\sigma}e^{-\frac{\omega^2}{2\sigma^2}}$, the $p$-th derivative of the Gaussian can be expressed as
\begin{equation}
    G_{\sigma}^{(p)}(\omega)=\frac{(-1)^p}{\sigma^p}G_{\sigma}(\omega)\mathrm{He}_{p}\!\left(\frac{\omega}{\sigma}\right),
    \tag{S14}
\end{equation}
where $\mathrm{He}_{p}(x)$ is the probabilists' Hermite polynomial, defined as $\mathrm{He}_{p}(x)=(-1)^p e^{x^2/2}\frac{d^p}{dx^p}e^{-x^2/2}$. Substituting into Eq.~\ref{eq:pu_ju}, we obtain
\begin{equation}\label{eq:pu_eq}
    A(k,\omega)=\sum_{p=0}^{\infty} 
                   \frac{\mu_{p}(k)}{p!\,\sigma^{p}}\,
                   G_{\sigma}(\omega)\mathrm{He}_{p}\!\left(\frac{\omega}{\sigma}\right).
                   \tag{S15}
\end{equation}

\section{VI. Convergence Analysis of Higher-Order Moments}

Equation~\ref{eq:pu_eq} can be decomposed into the contribution of each individual moment order:
\begin{equation}
    A(k,\omega)=\sum_{p=0}^{\infty}A_p(k,\omega),
    \tag{S16}
\end{equation}
where
\begin{equation}
    A_p(k,\omega)=\frac{\mu_{p}(k)}{p!\,\sigma^{p}}\,G_{\sigma}(\omega)\mathrm{He}_{p}\!\left(\frac{\omega}{\sigma}\right).
    \tag{S17}
\end{equation}

\begin{figure}[H]
    \centering
    \includegraphics[width=\linewidth]{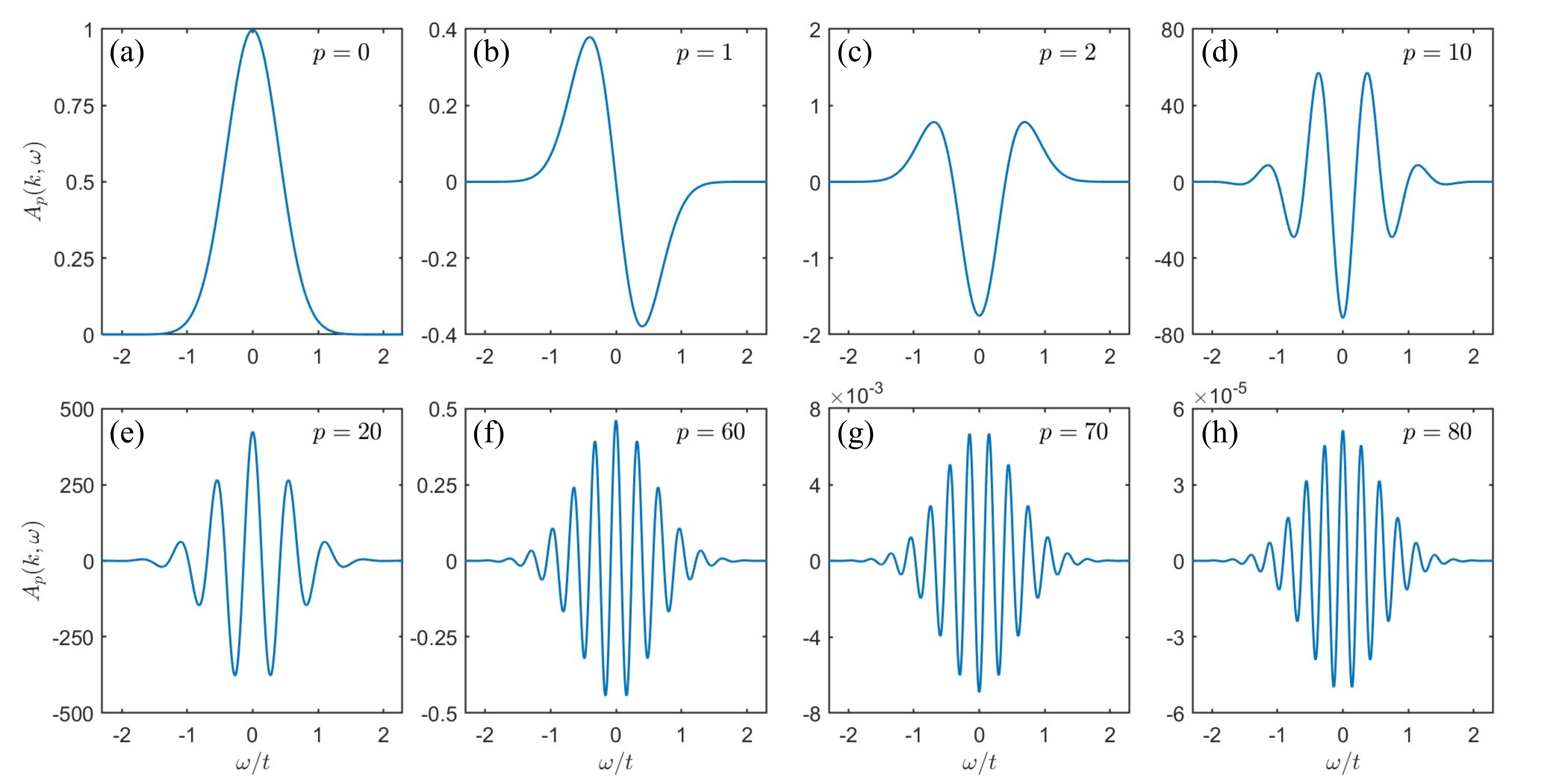}
    \caption{Contribution of each moment order to the spectral function at $V=t$ and $k=0.23G_1$.}
    \label{fig:puju2}
\end{figure}

\begin{figure}[H]
    \centering
    \includegraphics[width=0.5\linewidth]{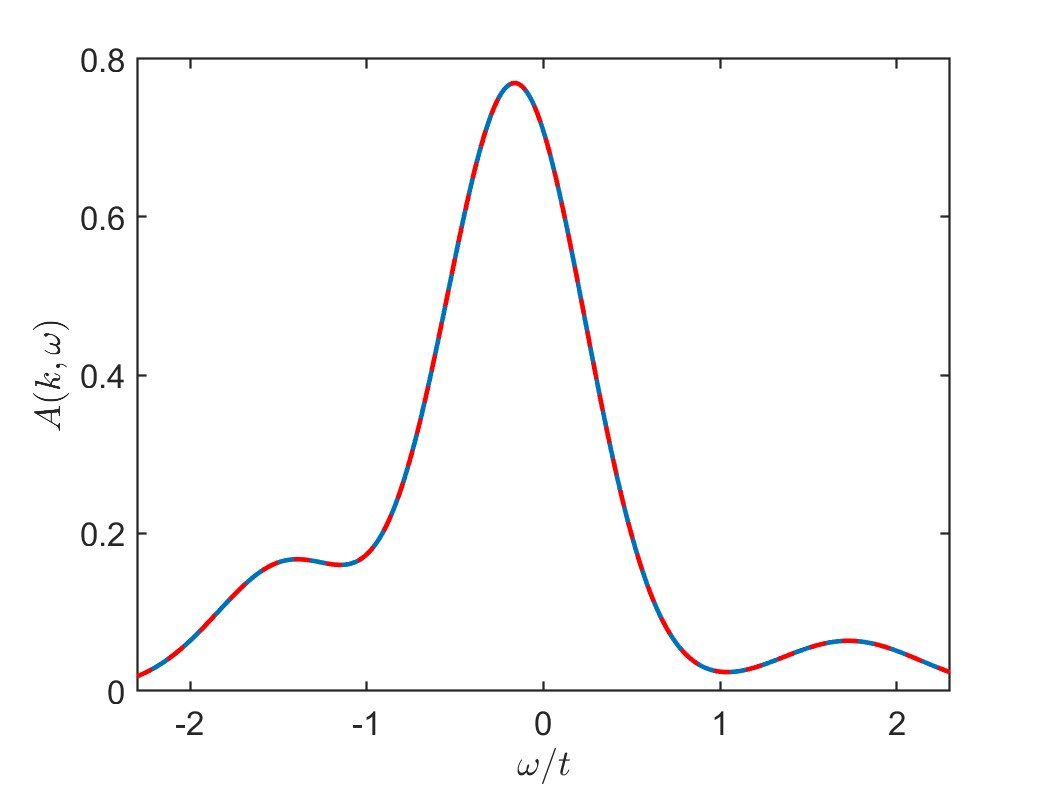}
    \caption{Spectral function obtained by matrix diagonalization (red) and by moment construction (blue dashed), with broadening uniformly set to $\sigma=0.4t$.}
    \label{fig:puju1}
\end{figure}

As $p$ increases, the contribution of higher-order terms first grows and then tends to zero. To make the demonstration visually apparent, a relatively large broadening $\sigma=0.4t$ is used. Fig.~\ref{fig:puju2} shows the contribution from each moment order; the higher-order terms are clearly very small, confirming the convergence of the truncation. In Fig.~\ref{fig:puju1}, we compare the spectral function obtained by diagonalization of the truncated Hamiltonian [Eq.~\eqref{Afunction}, red line] with that reconstructed from the moment expansion [Eq.~\eqref{eq:spectral-Gauss}, blue dashed]; the two coincide exactly.


\section{VII. Proof of the Moment-Locking Equality}

Let $Q(\omega)=\sum_{p=0}^M\alpha_p\omega^p$ be a polynomial of degree $M$. According to the moment-locking mechanism discussed in the main text, the first $N$ moments of the truncated Hamiltonian matrix of total size $N$ are exactly equal to the first $N$ moments of the true infinite Hamiltonian matrix; i.e., for $p\leq M\leq N$, $\mu_{p,c}(k)=\mu_p(k)$:
\begin{equation}
\begin{aligned}
    \int Q(\omega)A_c(k,\omega)d\omega&=\sum_{p=0}^M\alpha_p\mu_{p,c}(k)\\
    &=\sum_{p=0}^M\alpha_p\mu_p(k)\\
    &=\int Q(\omega)A(k,\omega)d\omega.
\end{aligned}
\tag{S18}
\end{equation}

For example, the average energy can be written as
\begin{equation}
\begin{aligned}
    \overline{E}&=\frac{1}{N_k}\sum_{k\in\mathrm{PBZ}}\int \omega A(k,\omega)d\omega\\
    &=\frac{1}{N_k}\sum_{k\in\mathrm{PBZ}}\mu_{1,c}(k).
\end{aligned}
\tag{S19}
\end{equation}

\section{VIII. Definition of the IPR}

The IPR is defined to characterize the localization properties of an eigenstate:
\begin{equation}
    \mathrm{IPR}(\ket{\psi})=\sum_{k\in PBZ}|\braket{k|\psi}|^4.
    \tag{S20}
\end{equation}

The energy-resolved IPR is defined as
\begin{equation}
\begin{aligned}
    \mathrm{IPR}(E)&=\frac{\sum_{k\in PBZ}\sum_m|\braket{k|\psi^{(m)}_k}|^4\delta(E-E^{(m)}_k)}{\sum_{k\in PBZ}\sum_m|\braket{k|\psi^{(m)}_k}|^2\delta(E-E^{(m)}_k)}.
\end{aligned}
\tag{S21}
\end{equation}

\section{IX. Definition of the LSEB}

Applying the duality transformation $\ket{r}=\frac{1}{\sqrt{N}}\sum_m e^{-iK_m\cdot r}\ket{K_m}$ to Eq.~\eqref{AAH_k_space_Lattice}, we obtain the real-space Hamiltonian\cite{chenTheoryLocalizedStates2025}:
\begin{equation}
    H_d=\sum_{r\in \mathrm{RBZ}}(V_rc^{\dagger}_rc_r+tc^{\dagger}_rc_{r\pm a_1}),
    \tag{S22}
\end{equation}
where $V_r=V\cos(G_2\cdot r+\theta)$. Here $c_r$ is the electron annihilation operator in real space, RBZ is the real-space Brillouin zone $[0,a_2)$, and $r$ is a continuous parameter on the RBZ. Define the real-space modulo lattice-point set $Q_{r}=\{R_m\mid R_m=(r+ma_1)\bmod a_2,m\in \mathbb{Z}\}$. By the same reasoning, we can write down the $r$-labeled Hamiltonian $H(r)$:
\begin{equation}
    H(r)=\sum_{m}(V_mc^{\dagger}_mc_m+tc^{\dagger}_mc_{m\pm 1}).
    \tag{S23}
\end{equation}
This expression is structurally identical to Eq.~\eqref{AAH_k_space_Lattice}, manifesting the duality property. When $V>2t$, the wave function is extended in $k$-space but localized in real space; we may therefore transplant the entire preceding theory and define the \emph{localized-state energy band} (LSEB) $E(r)=E^{(0)}_r$.

\section{X. $k$--$r$ Mapping Relation and Janus Quantum Numbers}

Owing to the duality relation, the real-space lattice index $m$ and the $k$-space lattice index $m$ are in one-to-one correspondence. The lowest-energy state in $k$-space is $\psi^{(0)}_{k=0}$, and the lowest-energy state in real space is $\psi^{(0)}_{r=\frac{a_2}{2}}$. Since this energy is non-degenerate, these two eigenstates must be one and the same eigenstate. An affine transformation between the PBZ and the RBZ can thus be established:
\begin{equation}
    r=a_2\!\left(\frac{1}{2}+\frac{k}{G_1}\right).
    \tag{S24}
\end{equation}
Under this mapping, the $k$-space lattice sites $K_m=mG_2\,(\bmod \,G_1)$ and the real-space lattice sites $R_m=\frac{a_2}{2}+ma_1\,(\bmod \,a_2)$ are in one-to-one correspondence, sharing the common lattice index $m$; in Fig.~4 of the main text, $m$ takes the values $0,1,\dots,6$. When $V<2t$, the wave function is localized in $k$-space and $m$ serves as the quantum number labeling $k$-space eigenstates; when $V>2t$, the wave function is localized in real space and $m$ serves as the quantum number labeling real-space eigenstates. We therefore call it a Janus quantum number. In the commensurate limit, translational symmetry is restored and the eigenstates become extended in real space; the Janus quantum number then reduces to the familiar Bloch momentum index $k$ and band index $n$.


\end{document}